\documentclass[amsmath,amssymb,amsfonts,aps,pre,preprint,superscriptaddress,bibnotes,showpacs,showkeys,longbibliography]{revtex4-1}

\usepackage{graphicx}
\usepackage{subfigure}
\usepackage{amsmath}
\usepackage{amsfonts}
\usepackage{amssymb}
\usepackage{natbib}
\usepackage{relsize}
\usepackage{sidecap}
\usepackage{braket}
\usepackage{footmisc}
\usepackage{footnote}
\usepackage{color}
\begin{document}

\title{Interfacial tension and wall energy of a Bose-Einstein condensate binary mixture: triple-parabola approximation}

\author{Zehui Deng}
\affiliation{Physics Department, Beijing Normal University, Beijing 100875, China}
\affiliation{Institute for Theoretical Physics, KU Leuven, Celestijnenlaan 200D,  BE-3001 Leuven, Belgium}

\author{Bert Van Schaeybroeck}
\affiliation{Royal Meteorological Institute, Ringlaan 3, BE-1180 Brussel, Belgium.}

\author{Chang-You Lin}
\affiliation{Institute for Theoretical Physics, KU Leuven, Celestijnenlaan 200D,  BE-3001 Leuven, Belgium}

\author{Nguyen Van Thu}
\affiliation{Department of Physics, Hanoi Pedagogical University 2, Hanoi, Vietnam}

\author{Joseph O. Indekeu}
\affiliation{Institute for Theoretical Physics, KU Leuven, Celestijnenlaan 200D,  BE-3001 Leuven, Belgium}

\begin{abstract}
Accurate and useful analytic approximations are developed for order parameter profiles and interfacial tensions of phase-separated binary mixtures of Bose-Einstein condensates. The pure condensates 1 and 2, each of which contains a particular species of atoms, feature healing lengths $\xi_1$ and $\xi_2$.  The inter-atomic interactions are repulsive. In particular, the effective inter-species repulsive interaction strength is $K$. A triple-parabola approximation (TPA) is proposed, to represent closely the energy density featured in Gross-Pitaevskii (GP) theory. This TPA allows us to define a model, which is a handy alternative to the full GP theory, while still possessing a simple analytic solution. The TPA offers a significant improvement over the recently introduced double-parabola approximation (DPA). In particular, a more accurate amplitude for the wall energy (of a single condensate) is derived and, importantly, a more correct expression for the interfacial tension (of two condensates) is obtained, which describes better its dependence on $K$ in the strong segregation regime, while also the interface profiles undergo a qualitative improvement. 
\end{abstract}

\maketitle

\section{Introduction}
At ultralow temperature phase separation of quantum gases has been realized experimentally, in particular in binary mixtures of Bose-Einstein condensates (BECs). For a survey, see, e.g., Refs.~\cite{UBFG,stamper,malomed2}. Early theoretical studies on trapped multicomponent BECs~\cite{ho,ejnisman,pu,timmermans,ao} appeared at the same time as experiments that witnessed weakly segregated dual BECs~\cite{modugno,miesner,myatt,stamper2,hall,matthews}. Ten more years were necessary to observe BEC components that are strongly segregated~\cite{papp} but by now these are commonly studied experimentally~\cite{mccarron,tojo,altin,xiong,mertes,nicklas,eto,wacker} by use of BECs with different spin states, isotopes and species. These experiments also revived the theoretical interest, much of which is focused on physical phenomena where the interface, separating the BEC components, plays a key role. Different contributions in this context can be found in~\cite{DPA1} and a few very recent and relevant research activities concern the statics of interface characteristics~\cite{IVS2,goldman,goldman2,polo}, capillary wave dispersion relations~\cite{takahashi,ticknor}, the kinetics~\cite{edmonds} and domain formation~\cite{eto,takeuchi1,nicklas} during phase segregation, and Nambu-Goldstone modes~\cite{takeuchi2,takahashi,roy1,roy2}.

In this paper we build further upon earlier work \cite{DPA1,ao,timmermans,BertVan,mazets,barankov2002boundary} devoted to the calculation of static interfacial properties of BEC binary mixtures within Gross-Pitaevskii (GP) theory \cite{GP}. Our main observation is that with a few new ideas and limited extra analytical work, a remarkable next step can be made towards physically transparent and powerful approximations that elucidate the physics of interfaces in soft condensed matter.

This paper is organized as follows. Section II recapitulates concisely the frame-work of GP theory. Section III defines the TPA model and illustrates its power in the context of the wall energy for a single condensate.  Section IV presents the application of the TPA to the interfacial tension and discusses our main analytical expression, with emphasis on the strong segregation regime.
Conclusions and an outlook are given in Section V.

\section{Brief recapitulation of Gross-Pitaevskii theory}
This section provides a summary of the theoretical frame-work given in more detail in Section II of \cite{DPA1}.
Consider the mean-field Gross-Pitaevskii Lagrangian density \cite{pethick2002bose,PitaString}
\begin{equation}\label{eq:lagrangiandensity}
\mathcal{L} (\Psi_1,\Psi_2 )=  \frac{i\hbar}{2}\sum_{j=1}^{2}  {\color{black} \left(\Psi_j^{\ast} \partial_t \Psi_j -\Psi_j \partial_t \Psi_j^{\ast} \right)} - \mathcal{E}(\Psi_1,\Psi_2),
\end{equation}
with Hamiltonian density
\begin{equation}\label{eq:hamiltoniandensity}
\mathcal{E} (\Psi_1,\Psi_2)=  \sum_{j=1}^{2} \left[ {\color{black} \frac{\hbar^2}{2 m_j}  \left|\nabla \Psi_j \right|^2 } + \frac{g_{jj}}{2}  \lvert\Psi_j\rvert^4  \right] + g_{12} \lvert\Psi_1\rvert^2 \lvert\Psi_2\rvert^2.
\end{equation}
For atomic species $j$, $\Psi_j=\Psi_{j}(\mathbf{r},t)$ is the wave function of the condensate or ``order parameter", $m_j$ the atomic mass, $g_{jj}=4 \pi \hbar^2 a_{jj}/m_j >0$ the repulsive intra-species interaction strength, $g_{12}=2 \pi \hbar^2 a_{12}(1/m_1 + 1/m_2)>0$ the repulsive inter-species interaction strength and $a_{jj'}$ the $s$-wave scattering length. 

By introducing the dimensionless quantities, $\mathbf{s}_j=\mathbf{r}/\xi_j$, with $\xi_j = \hbar/\sqrt{2m_jn_{j0}g_{jj}}$ the healing length and $n_{j0}$ the number density of condensate $j$ in bulk, $\tau_j=t/t_{j}$, $\psi_{j}=\Psi_{j}/\sqrt{n_{j0}}$, and $K=g_{12}/\sqrt{g_{11} g_{22}}$, where $t_{j}=\hbar/\mu_j$, and $\mu_j=g_{jj} n_{j0}$ the chemical potential of condensate $j$, we scale the Lagrangian density in \eqref{eq:lagrangiandensity} and Hamiltonian density in \eqref{eq:hamiltoniandensity} to
\begin{equation}\label{eq:dimensionlesslagrangiandensity}
\tilde{\mathcal{L}} (\psi_1,\psi_2) = \frac{\mathcal{L}}{2 P_0}=  \frac{i}{2} \sum_{j=1}^{2} {\color{black}\left(\psi_j^{\ast} \partial_{\tau_j} \psi_j  - \psi_j \partial_{\tau_j} \psi_j^{\ast} \right)} - \tilde{\mathcal{E}} (\psi_1,\psi_2),
\end{equation}
with
\begin{equation}
\tilde{\mathcal{E}} (\psi_1,\psi_2)= \frac{\mathcal{E}}{2 P_0} = \sum_{j=1}^{2} \left[ {\color{black} \left| \nabla_{\mathbf{s}_j} \psi_j \right|^2 }+ \frac{\lvert\psi_j\rvert^4}{2} \right] + K \lvert\psi_1\rvert^2 \lvert\psi_2\rvert^2,
\end{equation}
where the pressure $P_0$ is given by $\mu_j^2/2 g_{jj}$, which is independent of the label $j$ provided the mixture is at bulk two-phase coexistence. We restrict our attention in this paper to this two-phase equilibrium. 
Next we make a transformation of the dimensionless Lagrangian density by writing 
\begin{equation}
\psi_j (\mathbf{s}_j,\tau_j) \equiv \phi_j (\mathbf{s}_j,\tau_j)e^{- i\tau_j }.
\end{equation}
We then have a Lagrangian density in terms of the new order parameters $\phi_j$, 
\begin{equation}\label{eq:dimensionlesslagrangiandensityinphi}
\hat{\mathcal{L}} \left(\phi_1,\phi_2 \right) \equiv \tilde{\mathcal{L}} \left(\phi_1 e^{- i\tau_1}, \phi_2 e^{- i\tau_2} \right) = \sum_{j=1}^{2}  \left[  {\color{black} \frac{i}{2} \left( \phi_j^{\ast}  \partial_{\tau_j} \phi_j - \phi_j \partial_{\tau_j} \phi_j^{\ast} \right) - \left|\nabla_{\mathbf{s}_j} \phi_j \right|^2 } \right] - \hat{\mathcal{V}} (\phi_1,\phi_2),
\end{equation}
in which the potential $\hat{\mathcal{V}}$ takes the form
\begin{equation}\label{eq:potential}
\hat{\mathcal{V}}(\phi_1,\phi_2)= \sum_{j=1}^{2}  \left[-\lvert\phi_j\rvert^2 + \frac{\lvert\phi_j\rvert^4}{2} \right] + K \lvert\phi_1\rvert^2 \lvert\phi_2\rvert^2.
\end{equation}
Recall that for $K>1$ the two components are immiscible and a phase segregated BEC forms \cite{ao}.

It is well known that the Euler-Lagrange equations associated with $\hat{\mathcal{L}}$ yield
the time-dependent GP equations. These reduce 
 to the time-independent GP equations (TIGPE) when the order parameter $\phi_j $ is static, and these read
\begin{equation}\label{eq:TIGPEgeneralV}
 \nabla^2_{\mathbf{s}_j} \phi_{j} =   \frac{\partial \hat{\mathcal{V}}}{\partial \phi_{j}^{\ast} };\, j=1,2,
\end{equation}
which, for the given potential, implies
\begin{equation}\label{eq:TIGPE}
\left[- \nabla_{\mathbf{s}_j}^2 -1 + \lvert\phi_{j}\rvert^2 + K \lvert\phi_{j'}\rvert^2 \right] \phi_{j} = 0;  \;\; j=1,2 \; (j \ne j').
\end{equation}

For describing a planar interface at $z=0$, separating condensate 1 situated at $z \ge 0$ from condensate 2 situated at $z \le 0$, we consider order parameters that are translationally invariant in the $x$ and $y$ directions. For describing an interface the TIGPE must be solved with the boundary conditions
\begin{equation}\label{eq:planarbc}
\begin{split}
\phi_{1}(\rho_1\rightarrow \infty) &=\phi_{2}(\rho_2\rightarrow-\infty)=1 \\
\phi_{2}(\rho_2\rightarrow\infty) & =\phi_{1}(\rho_1\rightarrow-\infty) =0,
\end{split}
\end{equation}
where $\rho_j \equiv z/\xi_j$.

Likewise, 
for describing a condensate, say 1, adsorbed at an (ideal) optical wall \cite{IVS2} at $z=0$, we again consider an order parameter that is translationally invariant along $x$ and $y$, and solve the TIGPE with the boundary conditions
\begin{equation}\label{eq:planarbcwall}
\begin{split}
\phi_{1}(0) &= 0 \\
\phi_{1}(\rho_1\rightarrow \infty) &=1 \\
\end{split}
\end{equation}

Finally, we recall the limit of strong segregation $K \rightarrow \infty$, which plays an important role in our work. In this limit
the segregation  of the two condensates is complete. The overlap of the order parameters becomes zero and it does so in such a way that the interaction term $K \lvert\phi_1\rvert^2 \lvert\phi_2\rvert^2$ in the potential \eqref{eq:potential} becomes negligible. The GP equations decouple in this limit and the exact solution to the GP equations for the interface consists of two adjacent wall-like profiles \cite{Fetter}
\begin{equation}\label{eq:stationarysolutionsGPE}
\phi_{j} (\rho_j) = \tanh\left[ (-1)^{j+1}\frac{\rho_j}{\sqrt{2}}\right]; \; j=1,2.
\end{equation}

\section{Triple-Parabola Approximation (TPA) and wall energy}

Following up on the DPA introduced in \cite{DPA1} we expand to second order (harmonic approximation) the quartic potential $\hat{\mathcal{V}}$ in \eqref{eq:potential} about its two (equal) minima, which correspond to the bulk values for the order parameters. Moreover, and this defines the TPA, we {\em also} expand about its local maximum at the origin ($\phi_1=\phi_2=0$). However, the height of the maximum, $C_0$, is treated as an adjustable parameter, with respect to which the surface or interfacial energies can be minimized. 

In this Section we focus on a single-component BEC with order parameter $\phi_1$, and hence take $\phi_2 = 0$. For simplicity of notation, we set $\phi_1 = \phi$, $\xi_1 = \xi$ and $\rho_1=\rho$.
For obtaining the order parameter profile $\phi(\rho)$ in the half-space $z >0$, assuming a hard-wall boundary condition at $z=0$, we make use of, on the one hand, the expansion (to second order) of the GP potential
\begin{equation}\label{eq:GPpotentialSingle}
\hat{\mathcal{V}}(\phi, 0)= -\left|\phi\right|  ^2 + \frac{\left|\phi\right|  ^4}{2},
\end{equation}
about bulk condensate 1, which leads to \cite{DPA1}
\begin{equation}\label{eq:TPApotentialSinglePlus}
\hat{\mathcal{V}}_{\mathrm{TPA}}(\phi, 0)=   2\left(\left|\phi\right|  -1\right)^2  -\frac{1}{2},\;\mbox{for}\; \phi > \phi^{\times},
\end{equation}
and, on the other hand, the expansion about the origin in order parameter space, which leads to 
\begin{equation}\label{eq:TPApotentialSingleMinus}
\hat{\mathcal{V}}_{\mathrm{TPA}}(\phi, 0) = -\left|\phi\right|  ^2  + C_0,\;\mbox{for}\; \phi < \phi^{\times},
\end{equation}
where $\phi^{\times}$ is a matching value that is to be determined and $C_0$ is the adjustable height already discussed. Note that, for $\phi > \phi^{\times}$, we have $\hat{\mathcal{V}}_{\mathrm{TPA}}(\phi, 0)= \hat{\mathcal{V}}_{\mathrm{DPA}}(\phi, 0)$ \cite{DPA1}. Furthermore, we require that $\hat{\mathcal{V}}_{\mathrm{TPA}}(\phi, 0)$ be {\em continuous} at $\phi^{\times}$. With this requirement, \eqref{eq:TPApotentialSinglePlus} and \eqref{eq:TPApotentialSingleMinus} define the TPA {\em model} potential. Incidentally, it suffices for our purposes to work with a real order parameter, so the modulus signs can henceforth be dropped.

Fig.1 shows the functions that constitute the three potentials, GP, DPA and TPA. Note that for a certain range of ``shifts" $C_0$, there are {\em two} possible matching points for the TPA, where one can pass continuously from the inner (TPA) to the outer (DPA) branch (Fig.1a). For $C_0=1/6$ these two points merge into a single matching point, at which the branches are tangential, and $\phi^{\times}=2/3$ (Fig.1b). Note that, as compared to the DPA potential, the TPA potential is much closer to the GP potential.   

\setcounter{figure}{0}
\makeatletter 
\renewcommand{\thefigure}{\arabic{figure}a}
\begin{figure}
\begin{center}
\includegraphics[width=0.80\textwidth]{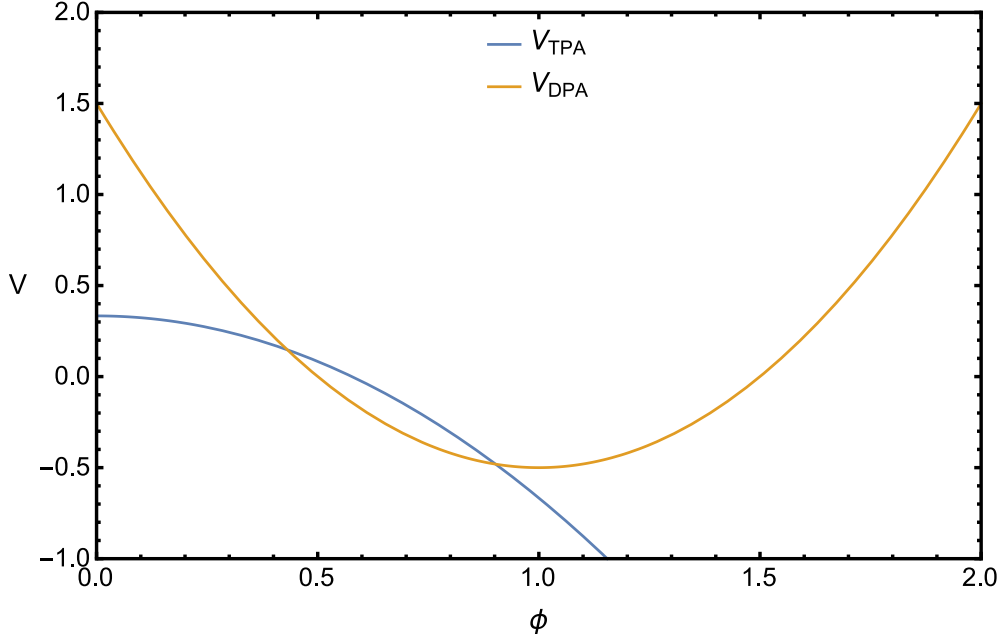}
\caption{\label{fig1a} (color online) Shown are the DPA potential (orange; convex curve) and the TPA branch (blue; concave curve) for the choice $C_0 = 1/3$ for the shift. For this choice two intersections appear. A possible TPA potential then consists of the TPA branch from $\phi = 0$ to either one of the two intersections and then switches to the DPA branch. 
   }     
\end{center}
\end{figure}
\setcounter{figure}{0}
\makeatletter 
\renewcommand{\thefigure}{\arabic{figure}b}
\begin{figure}
\begin{center}
\includegraphics[width=0.80\textwidth]{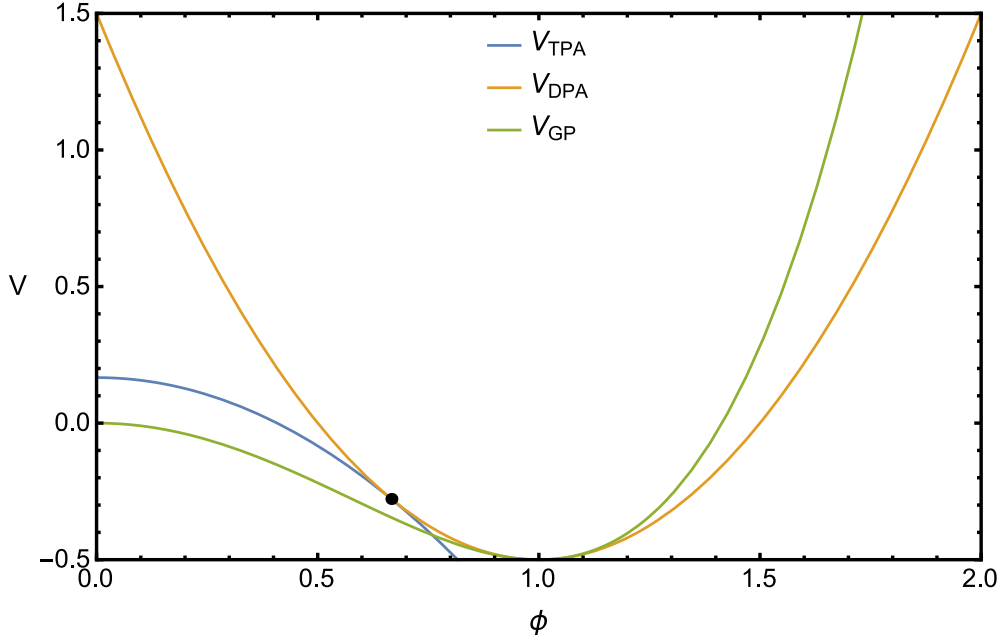}
\caption{\label{fig1b} (color online) Shown are the DPA potential (orange; convex curve), the TPA branch (blue; concave curve) and the GP potential (green) for the special choice $C_0 = 1/6$, for which the TPA branch is tangent to the DPA curve in the degenerate intersection point (black dot) at $\phi = 2/3$. The TPA potential follows the TPA branch from $\phi = 0$ to $\phi = 2/3$ and then switches (smoothly) to the DPA potential. 
   }     
\end{center}
\end{figure}

Using the TPA potential(s) and solving the EL equations \eqref{eq:TIGPEgeneralV} with the boundary conditions \eqref{eq:planarbcwall} is a simple exercise, which leads to the following order parameter profile, which is physically acceptable provided $ 1/6 \leq C_0 <  3/2$, i.e., $0<\phi^{\times}<1$,
\begin{eqnarray}\label{TPAs}
\phi(\rho)& = &\frac{\sqrt{2} \,\sin\rho}{\cos\lambda+\sqrt{2} \,\sin\lambda}, \;\mbox{for}\; \rho \leq \lambda \nonumber \\
\phi(\rho) &=& 1- \frac{\exp(-\sqrt{2} \,(\rho-\lambda))}{1+\sqrt{2} \,\tan\lambda}, \;\mbox{for}\; \rho \geq \lambda,
\end{eqnarray}
with $\lambda$ a matching point defined through $\phi^{\times} \equiv \phi(\lambda)$. The order parameter $\phi$ is, of course, required to be continuous at $\lambda$. Interestingly, the continuity of $\hat{\mathcal{V}}_{\mathrm{TPA}}$ at $\phi^{\times}$ implies that also the first derivative of the order parameter is continuous. This follows from insisting that a unique first integral of the EL equations, or ``constant of the motion" describes the entire trajectory (for more detail and a more general calculation, see Section IV). This is appropriate for equilibrium profiles, whereas for constrained (non-equilibrium) profiles, not considered in this work, the constant of the motion is in general discontinuous at the matching point \cite{IKHP}. 

There is still a free parameter in the problem, being $\lambda$, because any choice of $ 1/6 \leq C_0 <  3/2$ provides at least one (and at most two) values for $\lambda$. Interestingly, an optimal matching point $\lambda$ can be determined by minimizing the wall energy with respect to this parameter. The wall energy (or wall surface tension) is defined as the excess energy per unit area relative to the energy of a spatially uniform (bulk) order parameter. Within the TPA model the wall energy can be obtained as follows.

Starting from the grand potential for a condensate confined to the half-space $\rho>0$, 
\begin{equation}\label{eq:omega0wall}
\Omega_w[\phi,0]
 =2 P_0 \xi A \int_0^{\infty}  \mathrm{d} \rho  \left[ \left(\partial_{\rho} \phi \right)^2  + \hat{\mathcal{V}}_{\mathrm{TPA}} (\phi,0)\right],
\end{equation}
where $A$ is the wall area, and using the ``constant of the motion" \cite{DPA1},
\begin{equation}\label{eq:constantmotionwall}
 \left(\partial_{\rho} \phi \right)^2 - \hat{\mathcal{V}}_{\mathrm{TPA}} (\phi,0) = 1/2,
\end{equation}
we substitute, for order parameter profiles that satisfy the TIGPE, \eqref{eq:constantmotionwall} in \eqref{eq:omega0wall}, to obtain
the wall energy within the TPA model,
\begin{equation}\label{eq:wallenergy}
\gamma_{W1} = \frac{\Omega_w[\phi,0] + P_0V}{A} = 4 P_0 \xi \left ( \int_0^{\lambda}  + \int_{\lambda}^{\infty}\right ) \mathrm{d} \rho   \left(\partial_{\rho} \phi \right)^2.
\end{equation}
The integrals are readily performed and the result is the following function of $\lambda$,
\begin{equation}\label{eq:wallenergyresult}
\frac{\gamma_{W1}}{P_0\xi} = \frac{4(1+t^2)}{(1+\sqrt{2}\, t)^2}\,\lambda + \frac{2\sqrt{2}}{1+\sqrt{2}\, t},
\end{equation}
with $t\equiv \tan \lambda$. We can now optimize the parameter $\lambda$ by minimizing the wall energy with respect to it. This leads to the simple result
\begin{equation}\label{eq:wallenergyoptimal}
\tan \lambda = \sqrt{2}.
\end{equation}
This in turn implies a simple  matching value of the order parameter,
\begin{equation}\label{eq:orderparametermatching}
\phi^{\times}= \frac{2}{3}
\end{equation}
Remarkably, this minimum is achieved when the two branches of the TPA potential become tangent in a degenerate point of intersection (see Fig.1b). This happens for $C_0=1/6$. This tangency implies coincident derivatives, which in turn implies, through the EL equations, that also {\em the second derivative} of the order parameter is continuous at the matching point. The TPA order parameter profile is very smooth. We observe that, in a mechanical analogy of our model \cite{RowWid}, the order parameter is the position of a particle moving in a potential $V_{mech}= - \hat{\mathcal{V}}$, so that at the matching ``time" its position, its velocity and its acceleration are continuous (as is the force acting on it).

For this optimal matching point, the reduced wall energy takes the value
\begin{equation}\label{eq:wallenergyoptimalresult}
\frac{\gamma_{W1}^{(\rm{TPA})}}{P_0\xi} = \frac{4}{3}\,\tan^{-1}(\sqrt{2}) + \frac{2\sqrt{2}}{3} = 2.21656...,
\end{equation}
which is to be compared with the DPA approximation $2\sqrt{2}$ = 2.82842... \cite{DPA1}, and the exact value in GP theory, $4\sqrt{2}/3$ = 1.88561... 

Fig.2 allows one to compare qualitatively the order parameter profiles for a single condensate at a hard wall in the DPA and TPA models, against the exact solution of the full GP theory, which for this simple case is a tanh profile. Notice that the TPA leads to an overall improvement over the DPA, and is especially better near the wall, as expected, because the order parameter is small there. Its initial slope (0.816) is close to the exact one (GP theory; slope 0.707), whereas the initial slope predicted by the DPA model (1.414) is (much) larger. 

\setcounter{figure}{1}
\makeatletter 
\renewcommand{\thefigure}{\arabic{figure}}
\begin{figure}
\begin{center}
\includegraphics[width=0.80\textwidth]{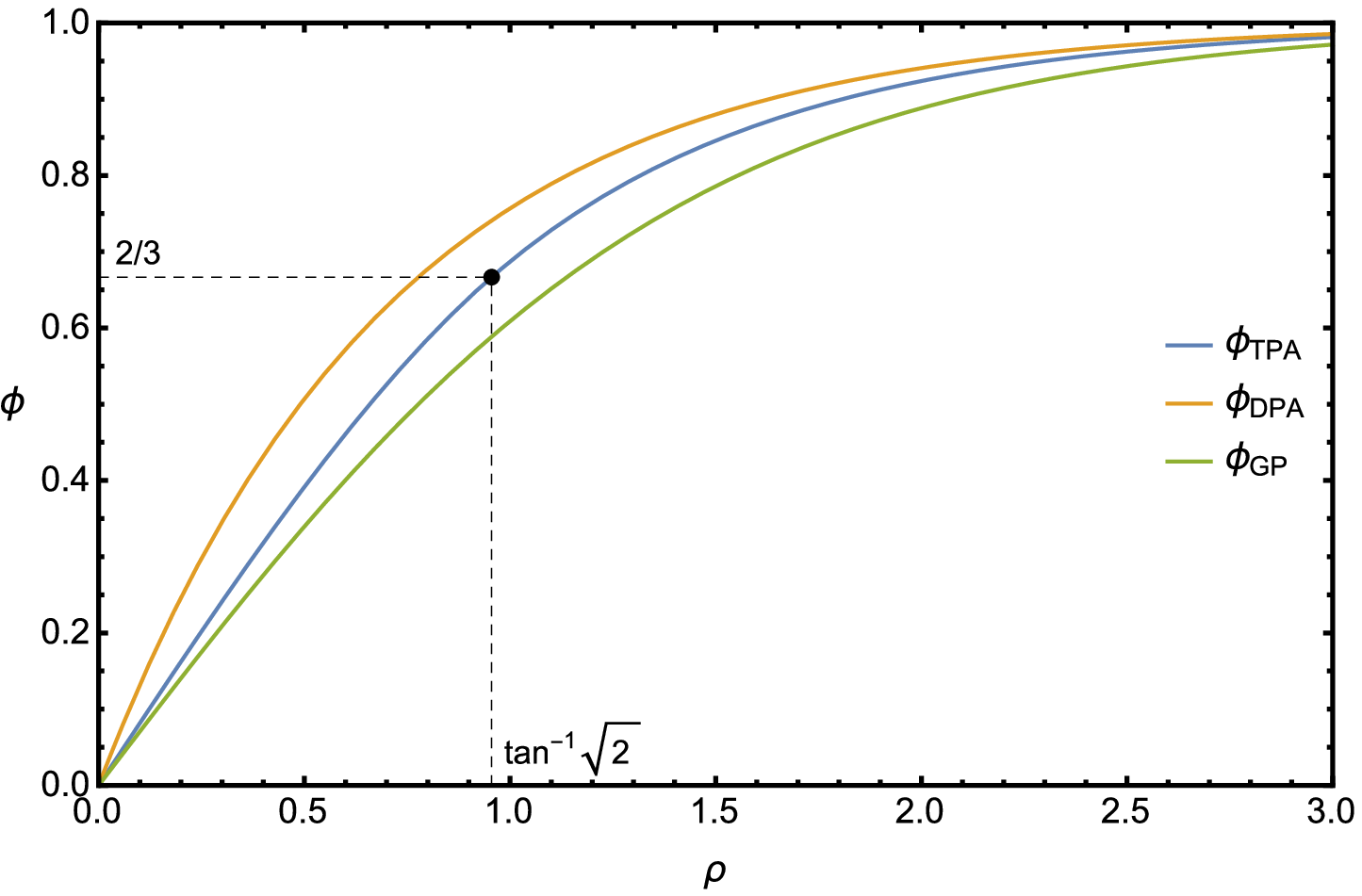}
\caption{\label{fig2}  (color online) Order parameter profiles for a single condensate at a hard wall. Shown are the (exact) GP profile (green) given by the analytic solution \eqref{eq:stationarysolutionsGPE} for $z>0$, the DPA (orange curve) and the TPA (blue curve) given by \eqref{TPAs}. The TPA profile features a matching point (black dot), at $(\tan^{-1} (\sqrt{2}), 2/3)$, corresponding to the degenerate intersection of the two potentials shown in Fig.1b. At this matching point the order parameter, its first and its second derivative are continuous. 
   }     
\end{center}
\end{figure}

\section{TPA and interfacial tension}

\subsection{TPA for the GP potential}

In this Section we apply the approach developed for a single condensate in the previous Section, to the binary mixture with order parameters $\phi_1 (\rho_1)$ and $\phi_2 (\rho_2) $. We aim at deriving the profiles for the interface between the two BECs and at calculating the interfacial tension.  
For obtaining the (real) order parameter profiles, subject to the boundary conditions \eqref{eq:planarbc}, we make use of, on the one hand, the expansions (to second order) of the GP potential \eqref{eq:potential}
about bulk condensates 1 and 2, for $z >0$ and $z<0$, respectively.  This leads to \cite{DPA1},
\begin{equation}\label{eq:TPApotentialDoublePlus}
\hat{\mathcal{V}}_{\mathrm{TPA}}(\phi_1,\phi_2) =  2\left(\left|\phi_j\right|  -1\right)^2 + (K-1) \lvert \phi_{j'} \rvert^2 -\frac{1}{2},
 \mbox{with} \;
\begin{cases}
z \ge 0, & (j,j')=(1,2),  \phi_1 \geq \phi_{+}^{\times},   \phi_2 \leq \phi_{-}^{\times}\\
z \le 0, & (j,j')=(2,1),  \phi_2 \geq \phi_{+}^{\times},  \phi_1 \leq \phi_{-}^{\times}
\end{cases}
\end{equation}
On the other hand, we also make use of the expansion about the origin in order parameter space, which leads to 
\begin{equation}\label{eq:TPApotentialDoubleMinus}
\hat{\mathcal{V}}_{\mathrm{TPA}}(\phi_1, \phi_2) = -\left|\phi_1\right|  ^2  -\left|\phi_2\right|  ^2+ C_0,\;\mbox{for}\; \phi_{-}^{\times} \leq \phi_1 \leq \phi_{+}^{\times},\;\phi_{-}^{\times} \leq \phi_2 \leq \phi_{+}^{\times},
\end{equation}
where $\phi_{+}^{\times}$ and $\phi_{-}^{\times}$ are two matching values that satisfy $\phi_{-}^{\times} \leq \phi_{+}^{\times}$ and are to be determined. Besides these three potential sheets, we need two more. Indeed, we also need ``mixed" potentials, to be used where one order parameter, but not the other, crosses over from one regime to the next. In this way we can guarantee continuity of $\phi_1$, $\phi_2$ and $\hat{\mathcal{V}}_{\mathrm{TPA}}$ everywhere, also when -- as is generally the case -- the matching points for $\phi_1$ and $\phi_2$ lie at different positions, i.e., different values of $z$. These mixed potentials read
\begin{equation}\label{eq:TPApotentialDoubleMixed}
\hat{\mathcal{V}}_{\mathrm{TPA}}(\phi_1,\phi_2) =  -\left|\phi_j\right|  ^2 + C_j + (K-1) \lvert \phi_{j'} \rvert^2,
\; \mbox{with} \;
\begin{cases}
z \ge 0, & (j,j')=(1,2),  \phi_1 \leq \phi_{+}^{\times},   \phi_2 \leq \phi_{-}^{\times}\\
z \le 0, & (j,j')=(2,1),  \phi_2 \leq \phi_{+}^{\times},  \phi_1 \leq \phi_{-}^{\times},
\end{cases}
\end{equation}
with the continuity and symmetry requirements $C_0 = C_1+C_2$ and $C_1=C_2$, respectively. 

Note that the potential sheet defined in \eqref{eq:TPApotentialDoublePlus} coincides with that of the DPA model \cite{DPA1}. Furthermore, we require that $\hat{\mathcal{V}}_{\mathrm{TPA}}(\phi_1, \phi_2)$ be {\em continuous} at $\phi_+^{\times}$ and $\phi_-^{\times}$ in both order parameters. With these requirements, \eqref{eq:TPApotentialDoublePlus}, \eqref{eq:TPApotentialDoubleMinus} and \eqref{eq:TPApotentialDoubleMixed} define the TPA {\em model} potential for the binary mixture. Incidentally, it suffices for our purposes to work with real order parameters, so the modulus signs can henceforth be dropped.

\subsection{TPA for the GP equations}

The triple-parabola-approximated GP equations are obtained by replacing the potential $\hat{\mathcal{V}}$ in the GP equations \eqref{eq:TIGPEgeneralV} by the potential $\hat{\mathcal{V}}_{\mathrm{TPA}}$, which leads to the following three sets of TIGPE equations for the TPA model,
 \begin{equation}\label{eq:TITPAGPEPlus}
\begin{split}
 \partial_{\rho_j}^2 \phi_{j} = {\color{black}2\left(\phi_{j}  -1\right) }   \\
 \partial_{\rho_{j'}}^2 \phi_{j'} = (K-1){\color{black}\phi_{j'} } 
\end{split} \; 
\; \mbox{with} \;
\begin{cases}
z \ge 0, & (j,j')=(1,2),  \phi_1 \geq \phi_{+}^{\times},   \phi_2 \leq \phi_{-}^{\times}\\
z \le 0, & (j,j')=(2,1),  \phi_2 \geq \phi_{+}^{\times},  \phi_1 \leq \phi_{-}^{\times}
\end{cases}
\end{equation}

\begin{equation}\label{eq:TITPAGPEMinus}
 \partial_{\rho_j}^2 \phi_{j} = - \phi_{j}     
\; \mbox{with} \; j =1,2; \; \phi_{-}^{\times} \leq \phi_1 \leq \phi_{+}^{\times},\;\phi_{-}^{\times} \leq \phi_2 \leq \phi_{+}^{\times}
\end{equation}

\begin{equation}\label{eq:TITPAGPEMixed}
\begin{split}
 \partial_{\rho_j}^2 \phi_{j} = -\phi_{j}     \\
 \partial_{\rho_{j'}}^2 \phi_{j'} = (K-1){\color{black}\phi_{j'} } 
\end{split} \; 
\; \mbox{with} \;
\begin{cases}
z \ge 0, & (j,j')=(1,2),  \phi_1 \leq \phi_{+}^{\times},   \phi_2 \leq \phi_{-}^{\times}\\
z \le 0, & (j,j')=(2,1),  \phi_2 \leq \phi_{+}^{\times},  \phi_1 \leq \phi_{-}^{\times},
\end{cases}
\end{equation}

\subsection{TPA for the constants of the motion}
It is instructive to give explicitly the first integrals of these equations of motion, the so-called constants of the motion. These are, in the respective regimes,
 \begin{equation}\label{eq:TITPACMPlus}
\begin{split}
 (\partial_{\rho_j} \phi_{j})^2 = 2\left(\phi_{j}  -1\right) ^2   \\
 (\partial_{\rho_{j'}} \phi_{j'})^2 = (K-1)\phi_{j'}^2 
\end{split} \; 
\; \mbox{with} \;
\begin{cases}
z \ge 0, & (j,j')=(1,2),  \phi_1 \geq \phi_{+}^{\times},   \phi_2 \leq \phi_{-}^{\times}\\
z \le 0, & (j,j')=(2,1),  \phi_2 \geq \phi_{+}^{\times},  \phi_1 \leq \phi_{-}^{\times}
\end{cases}
\end{equation}

\begin{equation}\label{eq:TITPACMMinus}
 (\partial_{\rho_j} \phi_{j})^2 = - \phi_{j}^2 + D_j     
\; \mbox{with} \; j =1,2; \; \phi_{-}^{\times} \leq \phi_1 \leq \phi_{+}^{\times},\;\phi_{-}^{\times} \leq \phi_2 \leq \phi_{+}^{\times}
\end{equation}

\begin{equation}\label{eq:TITPACMMixed}
\begin{split}
 (\partial_{\rho_j} \phi_{j})^2 = -\phi_{j}^2 + D_j    \\
 (\partial_{\rho_{j'}} \phi_{j'})^2 = (K-1)\phi_{j'}^2 
\end{split} \; 
\; \mbox{with} \;
\begin{cases}
z \ge 0, & (j,j')=(1,2),  \phi_1 \leq \phi_{+}^{\times},   \phi_2 \leq \phi_{-}^{\times}\\
z \le 0, & (j,j')=(2,1),  \phi_2 \leq \phi_{+}^{\times},  \phi_1 \leq \phi_{-}^{\times},
\end{cases}
\end{equation}
with, by symmetry, $D_1=D_2\equiv D$. These constants of the motion can be complemented by an equation which expresses the analogue of the conservation of total mechanical energy $E$. We have $E = T + V_{mech} = T - \hat{\mathcal{V}}_{\mathrm{TPA}}$, with $T$ the kinetic energy of the particle in the mechanical analogy. The value of $E$ is obtained as minus the value that $\hat{\mathcal{V}}_{\mathrm{TPA}}$ assumes in either one of the bulk phases, being minus $1/2$. We now assume, as is appropriate for equilibrium profiles (but not for non-equilibrium ones \cite{IKHP}), that one and the same constant $E$ applies to the entire trajectory. Then
the following useful identity holds, covering all the regimes of the TPA encountered in the half-space $z>0$ (the half-space $z<0$ is covered by interchanging the subscripts 1 and 2),
\begin{eqnarray}\label{eq:total energy}
E&=&(\partial_{\rho_1} \phi_{1})^2+(\partial_{\rho_2} \phi_2)^2+\phi_{1}^2+ \phi_{2}^2 - 2D + \frac{1}{2}\nonumber \\ &=& (\partial_{\rho_1} \phi_{1})^2+(\partial_{\rho_2} \phi_2)^2+\phi_{1}^2 - D + \frac{1}{2}  -(K-1)\phi_2^2\nonumber \\& =& (\partial_{\rho_1} \phi_{1})^2+(\partial_{\rho_2} \phi_2)^2-2(\phi_1-1)^2-(K-1)\phi_2^2+ \frac{1}{2} \nonumber\\&= &\frac{1}{2},
\end{eqnarray}
with $2D = C_0 + 1/2$, where $C_0$ has been introduced in \eqref{eq:TPApotentialDoubleMinus}.

\subsection{TPA solutions for the interface}
We now solve the TPA for the equations of motion in each regime and point out some properties of the solutions. For concreteness and to alleviate the notation we focus on the half-space $z \geq 0$. The solutions in the other half-space can be obtained by a judicious interchange of labels and/or signs, since the equations describing condensate 1 and 2 are the same when the scaled coordinates $\rho_1$ and $\rho_2$ are used. Close to the bulk phase (1,0) the solutions are the previously established DPA forms \cite{DPA1},
\begin{equation}\label{eq:stationarysolutionsDPAGPE}
\begin{split}
\phi_{1} (\rho_1) & = 1 - F_1 \,e^{- \sqrt{2}\, (\rho_1-\lambda_+) }, \; \mbox{for} \; \rho_1 \geq \lambda_+ \\
\phi_{2} (\rho_2) & = F_2 \,e^{- \sqrt{K-1} \,(\rho_2 - \lambda_- )}, \; \mbox{for} \; \rho_2 \geq \lambda_- 
\end{split}
\end{equation}
where, in line with our previous considerations, the matching points satisfy
$\phi_{+}^{\times}=\phi_1(\lambda_+)$ and  $\phi_{-}^{\times}=\phi_2(\lambda_-)$.

Close to the origin (0,0) in order parameter space, the solutions within the TPA are,
\begin{equation}\label{eq:stationarysolutionsDPAGPE}
\begin{split}
\phi_{1} (\rho_1) & = A \sin\rho_1 + \phi(0) \cos \rho_1, \; \mbox{for} \; \rho_1 \leq \lambda_+ \\
\phi_{2} (\rho_2) & = -A \sin\rho_2 + \phi(0) \cos \rho_2, \; \mbox{for} \; \rho_2 \leq \lambda_- ,
\end{split}
\end{equation}
where we have used the symmetry of the solutions and the interface midpoint property $\phi_1(0) = \phi_2(0) \equiv \phi(0)$. 

We now turn to simple arguments for determining the matching values 
$\phi_{+}^{\times}$ and $\phi_{-}^{\times}$ for the order parameters. From the constants of the motion we infer, for the matching of $\phi_1$,
\begin{equation}
-(\phi_{+}^{\times})^2 +D = 2(\phi_{+}^{\times} -1)^2,
\end{equation}
which has a degenerate solution $\phi_{+}^{\times} = 2/3$ at $D=2/3$, while for $D \gtrsim 2/3$ there are two solutions. We will be interested in the solution that minimizes the grand potential (see further). Next, for the matching of $\phi_2$ we obtain, again by using a constant of the motion, the simple result 
\begin{equation}
\phi_{-}^{\times}= \sqrt{\frac{D}{K}}.
\end{equation}

Expressing the continuity of the order parameters and their derivatives at the matching points provides us now with the following solutions for the amplitudes,
\begin{equation}
\phi(0) = \sqrt{D}\,\cos (\lambda_+ + \delta),
\end{equation}
\begin{equation}
A=\sqrt{D}\,\sin (\lambda_+ + \delta),
\end{equation}
with $\delta = \tan^{-1} \left(\sqrt{2}(1-\phi_{+}^{\times})/\phi_{+}^{\times}\right)$, and also
\begin{equation}
\phi(0) = \sqrt{D}\,\cos (\theta - \lambda_-),
\end{equation}
\begin{equation}
A=\sqrt{D}\,\sin (\theta - \lambda_-),
\end{equation}
with $\theta = \tan^{-1} (\sqrt{K-1})$,
so that
\begin{equation}\label{eq:sum}
\lambda_+ + \delta = \theta - \lambda_-,
\end{equation}
Furthermore,
\begin{equation}
F_1 = 1-\phi_{+}^{\times},
\end{equation}
\begin{equation}
F_2 = \sqrt{\frac{D}{K}}.
\end{equation}

The parameter $\lambda_-$ plays a special role. It is a free parameter in the model. Assuming $\lambda_- \geq 0$, from the TPA solutions it is easily seen that (for finite $K$) the second derivative of $\phi_2$ changes sign at $\rho_2 = \lambda_-$. Likewise the second derivative of $\phi_1$ changes sign at $\rho_1 = -\lambda_-$. For comparison, in the GP theory the second derivatives of the profiles pass (smoothly) through zero at points that are in general different from the midpoint of the interface (which lies at $z=0$). Exceptions are $K=3$ (solvable exactly \cite{Malomed}) and, evidently, $ K = \infty$, for which the two inflection points coincide with the midpoint (where the profiles intersect). In contrast, within the DPA the inflection points always coincide with the midpoint. The TPA is thus qualitatively different from the DPA in that it allows for the possibility that the inflections points differ from the midpoint. The TPA shares this property with the GP theory. However, the value of $\lambda_-$ is arbitrary. We shall see (in the next subsection) that, since the order parameters are decoupled, the interfacial tension within the TPA is {\em independent} of the choice of $\lambda_-$. Indeed, the grand potential only depends on the sum  $\lambda_+ + \lambda_-$. We consider it therefore most appropriate to choose  $\lambda_- $ such that {\em the interface midpoints of the DPA and TPA coincide}. This allows one to compare the interfacial profiles in the two models without bias. 

In view of these considerations we have,
\begin{equation}
\phi_{-}^{\times}= \sqrt{\frac{D}{K}},
\end{equation}
and
\begin{equation}\label{midpoint}
\phi (0)= \frac{\sqrt{2}}{\sqrt{2}+\sqrt{K-1}},
\end{equation}
where the latter is the result of the DPA \cite{DPA1}. We now proceed to study the interfacial tension variationally. This will allow us to obtain unique values for $D$, $\phi_{+}^{\times}$ and $\phi_{-}^{\times}$.

\subsection{TPA for the interfacial tension}
Using the constants of the motion we can calculate the interfacial tension within the TPA in terms of the following contributions,
\begin{equation}\label{eq:gamma12}
\gamma_{12} = \frac{\Omega-\Omega_{bulk}}{A}= 4 P_0 \sum_{j=1}^{2}\xi_j\int_{-\infty}^{\infty} \mathrm{d} \rho_j \left(\partial_{\rho_j} \phi_{j} \right)^2,
\end{equation}
which remain after the bulk grand potential $\Omega_{bulk}=-P_0 V$ of a homogeneous phase has been subtracted from the grand potential $\Omega$ of the configuration consisting of the two condensates and the interface in between them, i.e.,
\begin{equation}\label{eq:Omega}
\Omega[\phi_1,\phi_2]
 =2 P_0  A  \int_{-\infty}^{\infty}  \mathrm{d} z  \left[ \left(\partial_{\rho_1} \phi_1 \right)^2  + \left(\partial_{\rho_2} \phi_2 \right)^2 +\hat{\mathcal{V}}_{\mathrm{TPA}} (\phi_1,\phi_2)\right],
\end{equation}
Note that, at bulk two-phase coexistence,  $\Omega_{bulk}$ is the same for the two condensates.
Taking into account the different regimes encountered in the TPA and using the respective constants of the motion, we can rewrite \eqref{eq:gamma12} as
\begin{equation}\label{eq:gamma12regimes}
\frac{\gamma_{12}}{4 P_0} =  \xi_1\left (\int_{-\infty}^{-\lambda_-} + \int_{-\lambda_-}^{\lambda_+} + \int_{\lambda_+}^{\infty}\right)  \mathrm{d} \rho_1   \left(\partial_{\rho_1} \phi_{1} \right)^2 + \xi_2\left (\int_{-\infty}^{-\lambda_+} + \int_{-\lambda_+}^{\lambda_-} + \int_{\lambda_-}^{\infty}\right)  \mathrm{d} \rho_2   \left(\partial_{\rho_2} \phi_{2} \right)^2,
\end{equation}
Again using the constants of the motion this can be converted into integrals over $\phi_1$ and $\phi_2$, with the result
\begin{equation}\label{eq:gamma12result}
\frac{\gamma_{12}}{2 P_0} =  \left (D \left [ \sin^{-1}\left (\frac{\phi_{+}^{\times}}{\sqrt{D}} \right ) - \sin^{-1}\left (\frac{1}{\sqrt{K}} \right )  \right ] +\sqrt{2} (1-\phi_{+}^{\times}) \right )  (\xi_1+\xi_2),
\end{equation}
with $D= D(\phi_{+}^{\times}) = 3 (\phi_{+}^{\times})^2 - 4 \phi_{+}^{\times} +2$.
It is easily verified that $\gamma_{12}$ reaches a local minimum at $\phi_{+}^{\times} = 2/3$, precisely where $D$ has its minimum (and $D(2/3)= 2/3$). This local minimum {\em defines} the TPA. 

\setcounter{figure}{2}
\makeatletter 
\renewcommand{\thefigure}{\arabic{figure}}
\begin{figure}
\begin{center}
\includegraphics[width=0.65\textwidth]{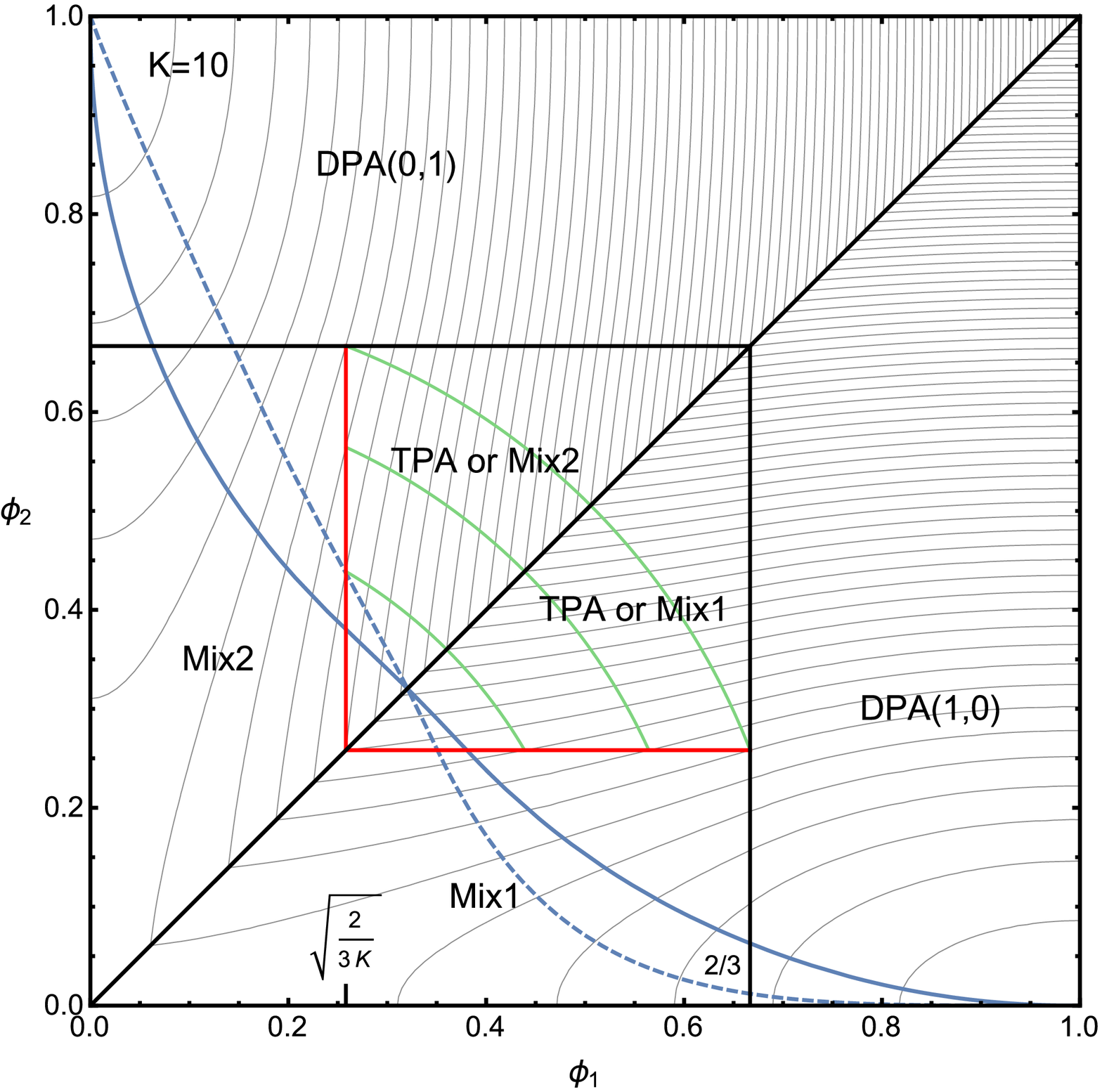}
\caption{\label{fig3}  (color online) Trajectory plots in the order parameter plane for $K=10$ and $\xi_1=\xi_2$ (symmetric case; solid blue line), and for $K=10$ and $\xi_1=2\xi_2$ (asymmetric case; dashed blue line). The trajectories represent an interface that connects the bulk phases at (1,0) and (0,1).  The contours of constant potential $\hat{\mathcal{V}}$ (grey lines) are shown for the potential sheets that are relevant in the different sectors visited by the trajectory. Close to the bulk points the potential is fully determined by the DPA appropriate for the bulk phase in the immediate vicinity of the trajectory, DPA(1,0) or DPA(0,1), and is given in \eqref{eq:TPApotentialDoublePlus}. When the trajectory traverses the black line to enter the square defined by $\phi_i < 2/3$, $i=1,2$, the relevant potential is of mixed character, Mix1 or Mix2, and is given in \eqref{eq:TPApotentialDoubleMixed}. Next, closer to the interface midpoint, when the trajectory enters the smallest square defined by $\sqrt{2/3K} < \phi_i < 2/3$, $i=1,2$, the relevant potential is of pure TPA character, as defined in \eqref{eq:TPApotentialDoubleMinus}, {\em provided} the trajectory crosses the red (South or West) borders of the square (as it does in the figure). The contour lines of the pure TPA potential sheet are segments of circles (green lines) and are rather far apart. For hypothetical trajectories that would enter the small square through the black (North or East) borders, the relevant potential remains the mixed one \eqref{eq:TPApotentialDoubleMixed}. The contour lines of the mixed potentials (grey) are more closely spaced.  Note that the relevant potentials are continuous across the (red or black) borders, as can be seen by inspecting the contours.  
}     
\end{center}
\end{figure}

We can now obtain the interface profiles as well as the interfacial tension uniquely. We first discuss the interface profiles. To this end we illustrate how the TPA works by showing examples of interface trajectories in the order parameter plane. Subsequently, we show the corresponding interface profiles themselves. Fig.3 presents the interface trajectory in the space of the two order parameters for $K=10$ (fairly strong segregation), for the symmetric case $\xi_1 = \xi_2$ and for the asymmetric case $\xi_1 = 2\xi_2$. Also shown in the figure are the contours of constant potential and the various sectors (triangles and squares) corresponding to different potential sheets encountered by the trajectory.  Fig.4 illustrates the interface trajectory for $K=3$ (intermediate segregation), for the symmetric case $\xi_1 = \xi_2$ and for the asymmetric case $\xi_1 = 2\xi_2$. Note that for the symmetric case the trajectory  coincides with the antidiagonal of the quadrant. This perfect symmetry ($\phi_2 = 1-\phi_1$) is a property which the DPA shares with the exact solution of the GP equations for $K=3$ \cite{Malomed}.

\setcounter{figure}{3}
\makeatletter 
\renewcommand{\thefigure}{\arabic{figure}}
\begin{figure}
\begin{center}
\includegraphics[width=0.70\textwidth]{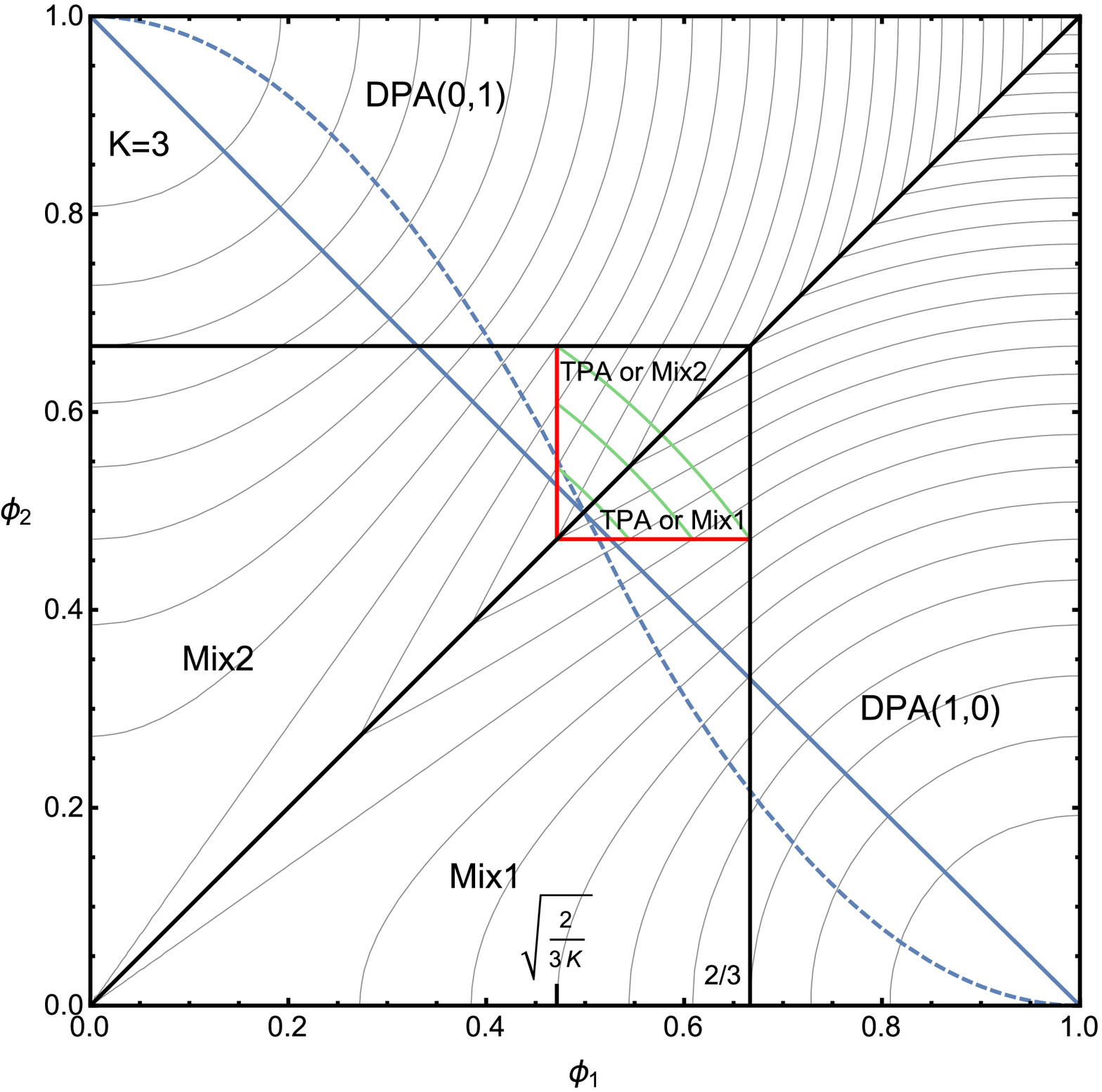}
\caption{\label{fig4}  (color online) Trajectory plots in the order parameter plane for $K=3$ and $\xi_1=\xi_2$ (symmetric case; solid blue line), and  for $K=3$ and $\xi_1=2\xi_2$ (asymmetric case; dashed blue line). The description of the figure is similar to that of Fig.3 and we refer to that figure for details.
   }     
\end{center}
\end{figure}

Fig.5a shows the interface profiles corresponding to $K=10$ and $\xi_1 = \xi_2$, while Fig.5b presents the interface profiles corresponding to $K=10$ and $\xi_1 = 2\xi_2$. In each figure, three profiles are displayed: the numerical solution to the GP equations, the DPA and the TPA. As we described, the TPA profiles are characterized by weak singularities at the matching points (black dots). At the upper matching points the order parameter and its first and second derivatives are continuous, whereas at the lower matching points the second derivative is discontinuous. Note that the TPA offers a modest improvement over the DPA. Recall that, by construction, the interface midpoints within the DPA and the TPA coincide. Likewise, Fig.6a shows the interface profiles corresponding to $K=3$ and $\xi_1 = \xi_2$, while Fig.6b presents the interface profiles corresponding to $K=3$ and $\xi_1 = 2\xi_2$. For this case of intermediate segregation the DPA and the TPA are almost equivalent. Recall that the exact solution to the GP equations, for $K=3$ and $\xi_1 = \xi_2$, takes a very simple tanh form \cite{Malomed} and that the midpoint lies at $\phi = 1/2$, a property which is also respected by the DPA (and TPA). 

\setcounter{figure}{4}
\makeatletter 
\renewcommand{\thefigure}{\arabic{figure}a}
\begin{figure}
\begin{center}
\includegraphics[width=0.80\textwidth]{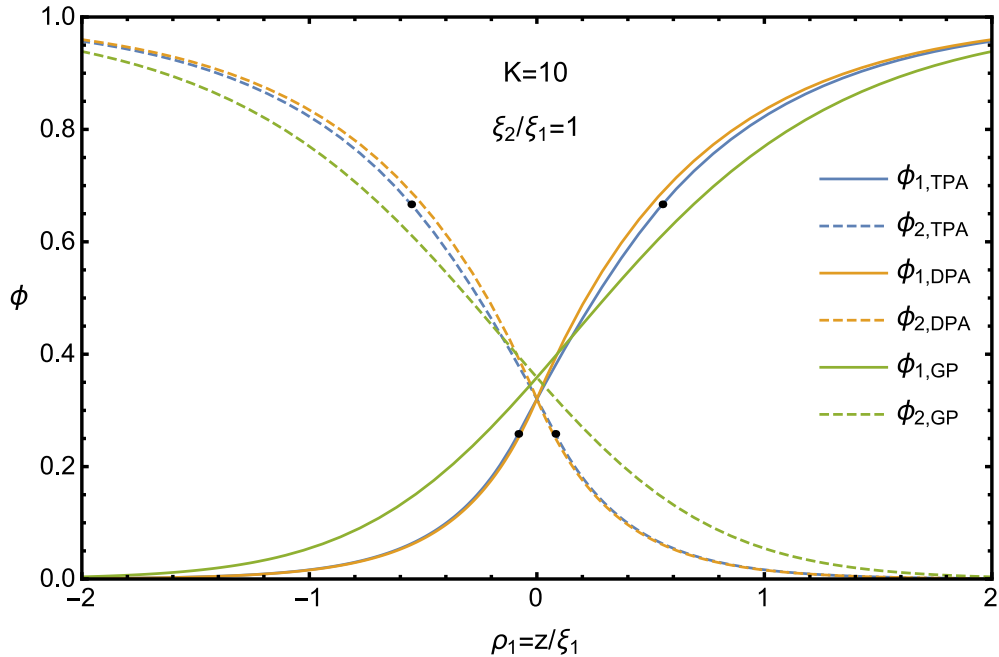}
\caption{\label{fig5a}  (color online) Interface profiles for $K=10$ and $\xi_1 = \xi_2$ (symmetric case). The numerically exact solution within GP theory is shown (solid green and dashed green lines), together with the DPA solutions (solid orange and dashed orange lines) and the TPA solutions (solid blue and dashed blue lines). The latter feature (weak) singularities at the matching points.
   }     
\end{center}
\end{figure}
\setcounter{figure}{4}
\makeatletter 
\renewcommand{\thefigure}{\arabic{figure}b}
\begin{figure}
\begin{center}
\includegraphics[width=0.80\textwidth]{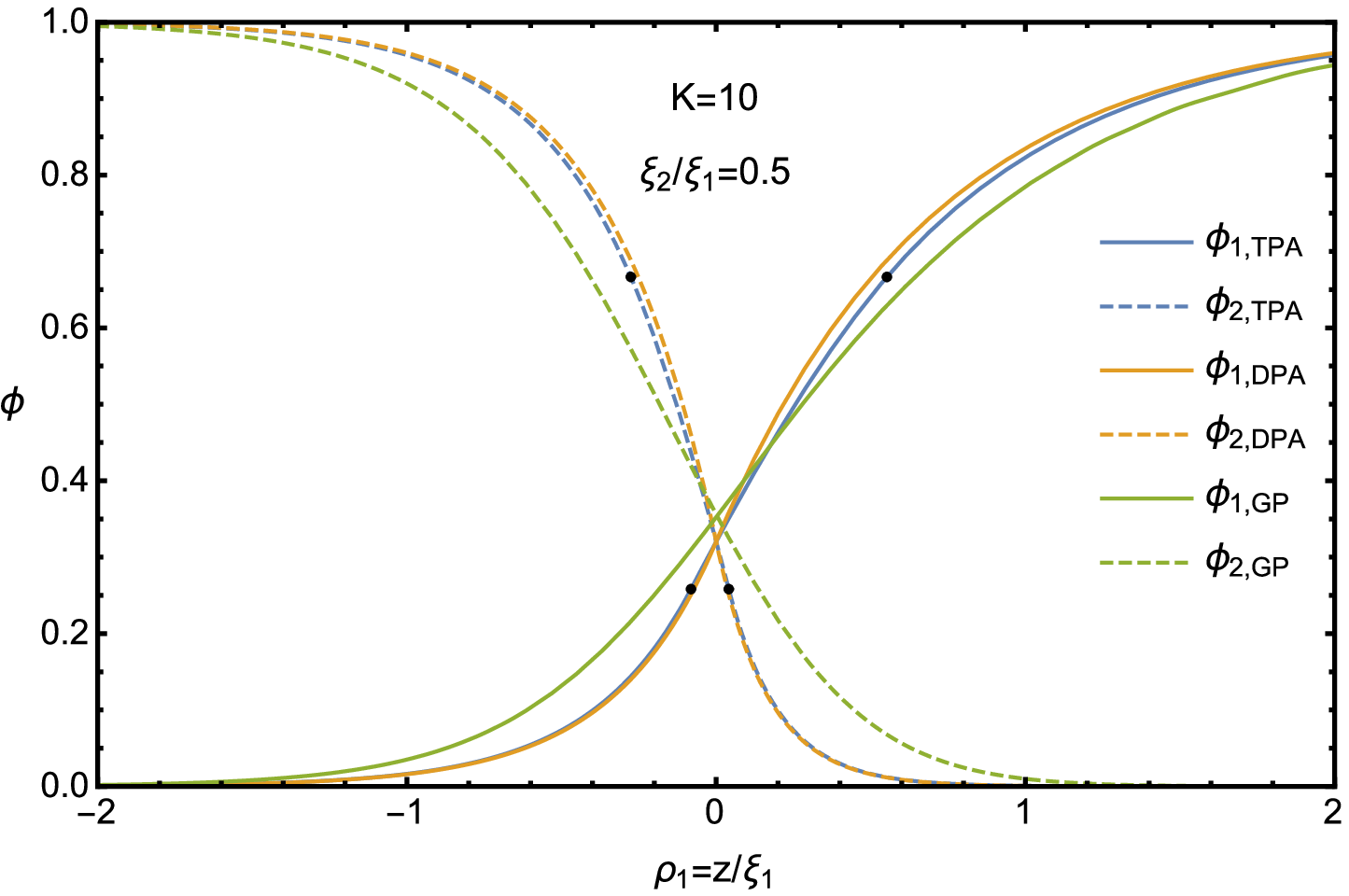}
\caption{\label{fig5b}  (color online) Interface profiles for $K=10$ and $\xi_1 = 2\xi_2$ (an asymmetric case). 
   }     
\end{center}
\end{figure}
\setcounter{figure}{5}
\makeatletter 
\renewcommand{\thefigure}{\arabic{figure}a}
\begin{figure}
\begin{center}
\includegraphics[width=0.80\textwidth]{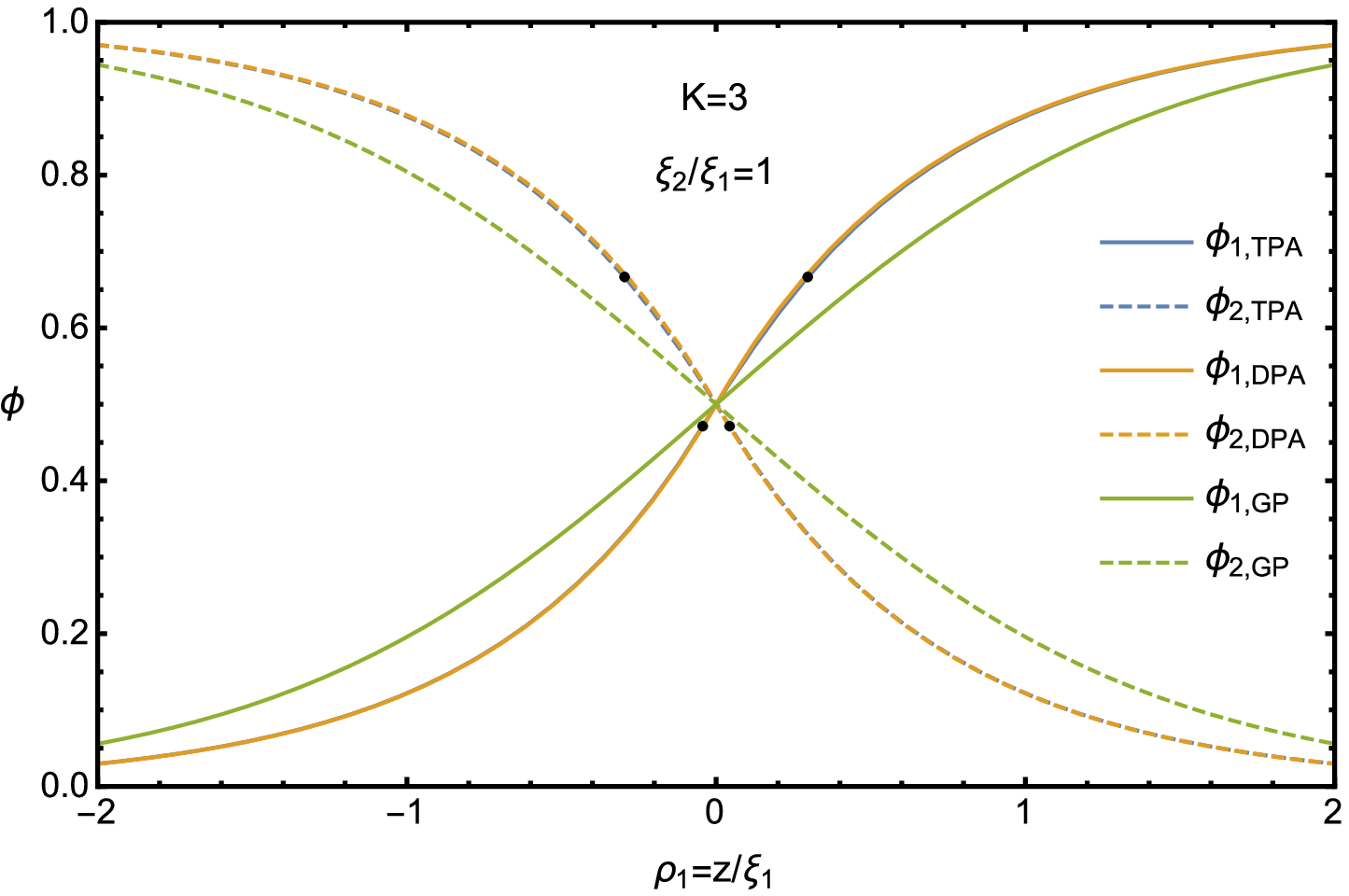}
\caption{\label{fig6a}  (color online) Interface profiles for $K=3$ and $\xi_1 = \xi_2$ (symmetric case). The analytically known exact solution within GP theory is shown (solid green and dashed green lines), together with the DPA solutions (solid orange and dashed orange lines) and the TPA solutions (solid blue and dashed blue lines). The latter feature (weak) singularities at the matching points.
   }     
\end{center}
\end{figure}
\setcounter{figure}{5}
\makeatletter 
\renewcommand{\thefigure}{\arabic{figure}b}
\begin{figure}
\begin{center}
\includegraphics[width=0.80\textwidth]{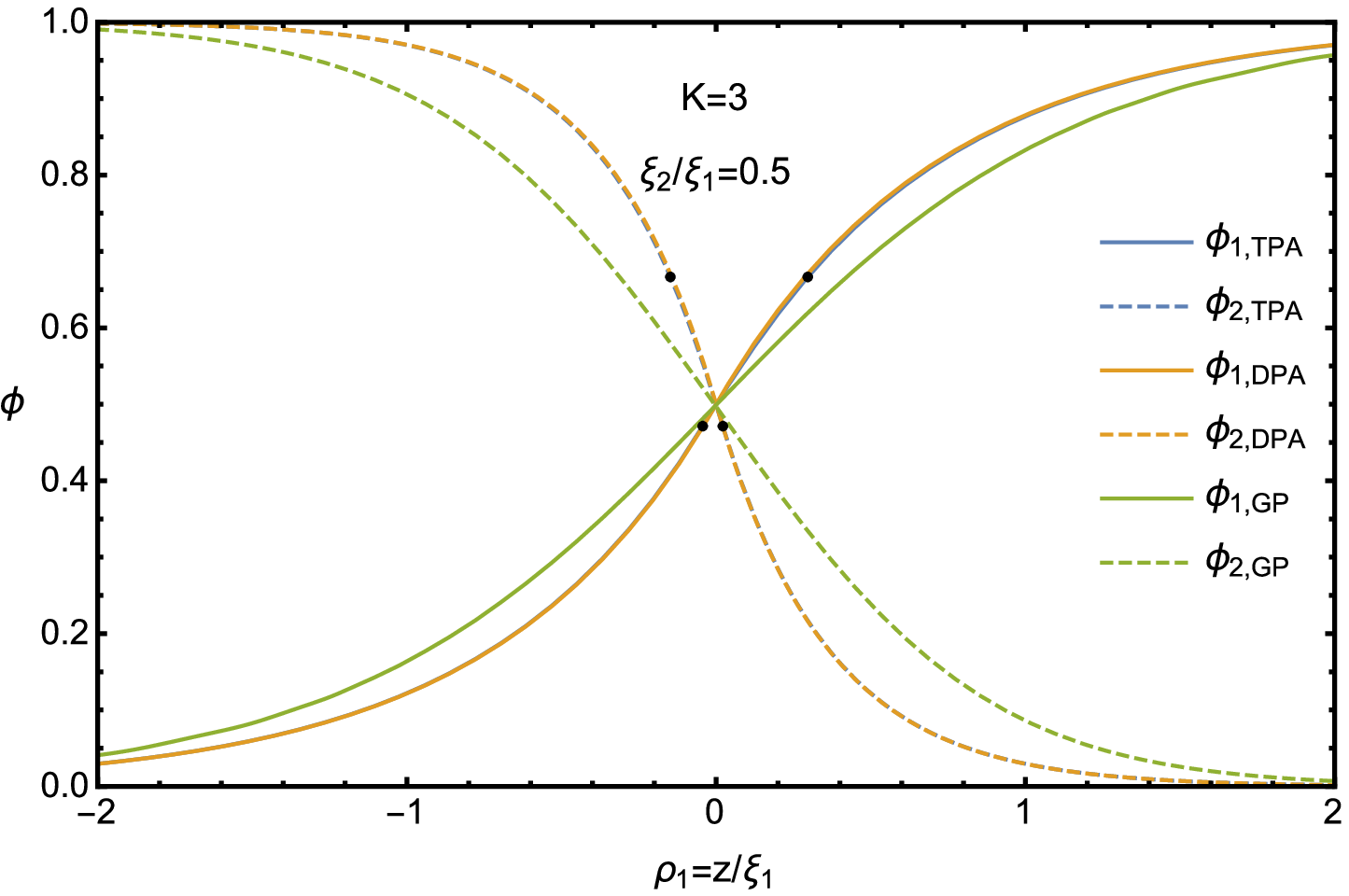}
\caption{\label{fig6b}  (color online) Interface profiles for $K=3$ and $\xi_1 = 2\xi_2$ (an asymmetric case). 
   }     
\end{center}
\end{figure}

Next we return to the interfacial tension. Substituting into \eqref{eq:gamma12result} the special parameter values we determined, 
we arrive at our main result
\begin{equation}\label{eq:gamma12resultTPA}
\gamma_{12}^{\rm (TPA)} = \frac{4}{3}  \left ( \sin^{-1}\left (\sqrt{\frac{2}{3}}\right ) - \sin^{-1}\left (\frac{1}{\sqrt{K}}\right )    + \frac{1}{\sqrt{2}} \right )  P_0 (\xi_1+\xi_2),
\end{equation}
This expression can be used for $K \geq 3/2$. Indeed, for $K \downarrow 3/2$ the interface midpoint value \eqref{midpoint} approaches the matching value $\phi_{+}^{\times} = 2/3$ and the matching position $\lambda_+$ approaches the midpoint position $z=0$. Fig.8 shows the interfacial tension versus $1/K$ in GP theory and in the  DPA and TPA models. 

We can obtain $\lambda_+ + \lambda_-$ explicitly as a function of $K$ using \eqref{eq:sum},
which leads to
\begin{equation}
\lambda_+ + \lambda_- = \tan^{-1} (\sqrt{K-1}) - \tan^{-1} \left (\frac{1}{\sqrt{2}}\right ) = \sin^{-1}\left (\sqrt{\frac{2}{3}}\right ) - \sin^{-1}\left (\frac{1}{\sqrt{K}}\right ).
\end{equation}
This function is displayed in Fig.7. We recall that only the sum $\lambda_+ + \lambda_-$ is relevant for the interfacial tension, while shifts of the individual matching point positions $\lambda_+$ and $\lambda_-$ affect the profiles but not the interfacial tension as long as their sum is invariant.

\setcounter{figure}{6}
\makeatletter 
\renewcommand{\thefigure}{\arabic{figure}}
\begin{figure}
\begin{center}
\includegraphics[width=0.80\textwidth]{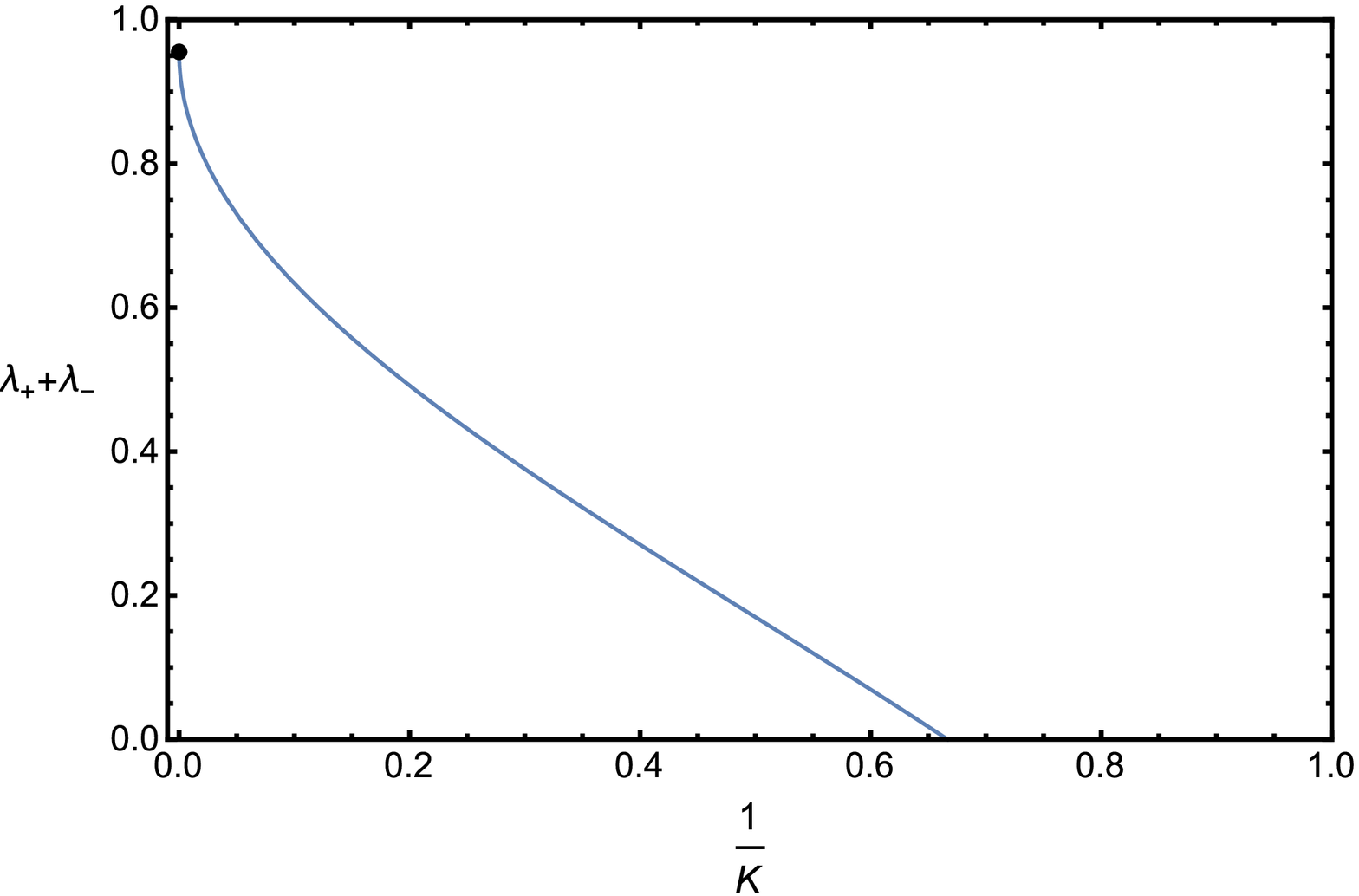}
\caption{\label{fig7}  (color online) Dependence of the matching point position sum $\lambda_+ + \lambda_-$ on the inverse of the interaction parameter $K$. For $1/K=0$ this sum takes the value $\tan^{-1} (\sqrt{2}) = 0.955...$ and it reaches zero at $K=3/2$, where the applicability of the  TPA ends. The individual matching point positions  $\lambda_+$ and $\lambda_-$ can both be determined numerically through the requirement that the interface midpoint value of the order parameters be identical in DPA and TPA.
   }     
\end{center}
\end{figure}

The vanishing of $\lambda_+ + \lambda-$ at $K=3/2$ signifies that in the interval $1< K \leq 3/2$ the TPA reduces simply to the DPA. Recall that $K>1$ is a necessary condition for phase segregation of the condensates. For $K > 3/2$ the TPA model becomes relevant and leads to an interfacial tension that is lower than that in the DPA model. Note that in the strong segregation limit ($K \rightarrow \infty$) we recover, as we should, the sum of two wall tensions,
\begin{equation}\label{eq:wallrecovery}
\gamma_{12}^{(\rm{TPA})} \rightarrow   \gamma_{W1}^{(\rm{TPA})} + \gamma_{W2}^{(\rm{TPA})},\; \mbox{for} \; K \rightarrow \infty,
\end{equation}
with the wall tension as obtained in \eqref{eq:wallenergyoptimalresult}.

For comparison, we recall the result for the interfacial tension in the DPA model \cite{DPA1},
\begin{equation}\label{eq:gamma12DPA}
\gamma_{12}^{(\rm{DPA})} =  2 \sqrt{2} \frac{\sqrt{(K-1)/2}}{1+\sqrt{(K-1)/2}}P_0 (\xi_1 + \xi_2).
\end{equation}

\setcounter{figure}{7}
\makeatletter 
\renewcommand{\thefigure}{\arabic{figure}}
\begin{figure}
\begin{center}
\includegraphics[width=0.80\textwidth]{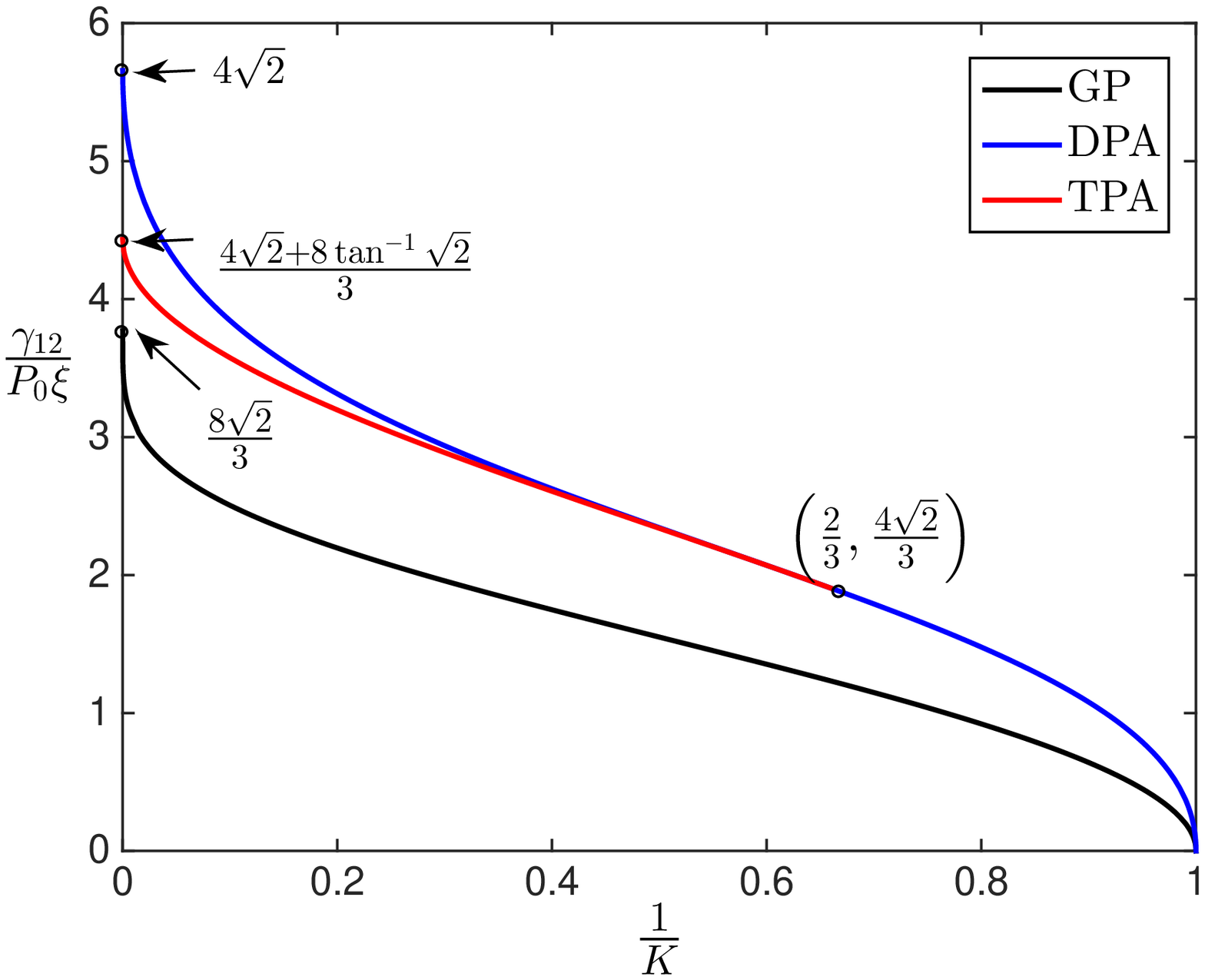}
\caption{\label{fig8}  (color online) Reduced interfacial tension versus the inverse of the interaction parameter $K$. Shown are the numerically exact result within GP theory (lower curve;  black), the DPA result (upper curve; blue) and the TPA refinement (middle curve; red) applicable for $0 < 1/K < 2/3$. The respective values in the strong segregation limit ($1/K=0$) are indicated at the arrows. Also indicated is the point where the TPA curve intersects the DPA curve at $K=3/2$ (with unequal slopes). For $K < 3/2$ (weak segregation regime) the interface trajectory does not visit the pure TPA sector (cf. Fig.3 or Fig.4).  
   }     
\end{center}
\end{figure}

\section{Conclusion}

In this work we have introduced and applied an analytic approach, the TPA, that offers a useful improvement over the well-known DPA strategy. In the TPA, harmonic approximations to a potential are not only invoked near the minima or fixed points, but also near a local maximum, in as far as the smooth curvature near this maximum, but not its height, are respected. This is a meaningful approach, because in contrast to the value in the minima (equilibrium bulk solutions), the value in the maximum is in itself not of physical significance. While the DPA is a widely applied approximation in the context of variational problems in particle mechanics and statistical mechanics, we are not aware of any previous study introducing or employing a TPA.  Quite generally, the TPA allows one to   improve over the DPA while still obtaining analytical results with modest calculational effort.

Applying the TPA to interfacial properties of mixtures of BECs, we have arrived at the following conclusions. 
\begin{itemize}
\item For the wall energy the TPA result constitutes a significant quantitative improvement over the DPA result. Moreover, the variational calculation intrinsic to the TPA leads to an order parameter profile with remarkable smoothness (at the matching point it is continuous and possesses continuous first and second derivatives). Furthermore, the initial slope of the order parameter (at the wall) within the TPA is also much improved with respect to the DPA. 
\item 
For the interfacial profiles and the interfacial tension the TPA offers a modest but interesting improvement over the DPA results. In the strong segregration regime the improvement for the interfacial tension is significant, simply because the wall energies are (much) more accurate in the TPA. In the weak segregation limit the TPA leads to the same interface structure and interfacial tension as the DPA, simply because TPA and DPA coincide for $1 < K < 3/2$. For intermediate and strong segregation, say $K>3$, the TPA becomes worthwhile and offers a significant improvement over the DPA for profiles and interfacial tension. Furthermore, the TPA shares the qualitatively important feature with the (exact) GP theory that the inflection points of the profiles need not coincide with the midpoint of the interface.
\item The analytic expression for the interfacial tension within TPA -- our main result \eqref{eq:gamma12resultTPA} -- is of a remarkable simplicity, and provides a useful and compact refinement of the DPA result, especially for strong segregation.
\end{itemize}

\begin{acknowledgments}
N.V.T is supported by the Vietnam National Foundation for Science and Technology Development (NAFOSTED)  J.O.I. and C.-Y.L. by FWO Flanders under Grant Nr. FWO.103.2013.09 within the framework of the FWO-NAFOSTED cooperation. J.O.I. and C.-Y.L. are furthermore supported by  KU Leuven Grant OT/11/063. N.V.T. is also funded by Ministry of Education and Training of Vietnam. The authors thank Phat Tran Huu, Xintian Wu, Wenan Guo and Jesper Koning for discussions.

\end{acknowledgments}


\end{document}